\documentclass[aps,showpacs,preprintnumbers,amsmath,amssymb]{revtex4}
\usepackage{dcolumn}
\usepackage[dvips]{epsfig}

\begin{document}
\baselineskip=0.8 cm
\title{\bf Dynamical evolution of a scalar field coupling to Einstein's tensor in the Reissner-Nordstr\"{o}m black hole spacetime}

\author{Songbai Chen\footnote{csb3752@163.com}, Jiliang Jing
\footnote{jljing@hunnu.edu.cn}}
%\email{csb3752@163.com}

\affiliation{Institute of Physics and Department of Physics, Hunan
Normal University,  Changsha, Hunan 410081, People's Republic of
China \\ Key Laboratory of Low Dimensional Quantum Structures \\
and Quantum Control of Ministry of Education, Hunan Normal
University, Changsha, Hunan 410081, People's Republic of China}

\begin{abstract}
\baselineskip=0.6 cm
\begin{center}
{\bf Abstract}
\end{center}

We study the dynamical evolution of a scalar field coupling to
Einstein's tensor in the background of Reissner-Nordstr\"{o}m black
hole. Our results show that the the coupling constant $\eta$
imprints in the wave dynamics of a scalar perturbation. In the weak
coupling, we find that with the increase of the coupling constant
$\eta$ the real parts of the fundamental quasinormal frequencies
decrease and the absolute values of imaginary parts increase for
fixed charge $q$ and multipole number $l$. In the strong coupling,
we find that for $l\neq0$ the instability occurs when $\eta$ is
larger than a certain threshold value $\eta_c$ which deceases with
the multipole number $l$ and charge $q$. However, for the lowest
$l=0$, we find that there does not exist such a threshold value and
the scalar field always decays for arbitrary coupling constant.

\end{abstract}

\pacs{ 04.70.Dy, 95.30.Sf, 97.60.Lf } \maketitle
\newpage
\section{Introduction}

The dynamical evolution of an external field perturbation around a
black hole has been an object of great interest for physicists for
the last few decades.  One of the main reasons is that the
frequencies of the quasinormal oscillations appeared in the
dynamical evolution carry the characteristic information about the
black hole, which could provide a way for astrophysicists to
identify whether there exists black hole in our Universe or not
\cite{Chandrasekhar:1975,Nollert:1999,Kokkotas:1999}. The further
studies indicate that the quasinormal spectrum could help us to
understand more deeply about the quantum gravity
\cite{Hod:1998,Dreyer:2003,Corichi:2003} and the AdS/CFT
correspondence \cite{Maldacena:1998,Witten:1998,Horowitz:2000}.
Moreover, the stability of a black hole can be examined by the study
of the dynamical behaviors of the perturbations in the background
spacetime \cite{Gregory:1993,Harmark:2007,Konoplya:2008,Chen:2009}.
Therefore, a lot of attention have been focused on the dynamical
evolution of various perturbations in the various black holes
spacetime.

The simplest field in the quantum field theory is scalar field,
which associated with spin-$0$ particles. The dynamical evolution of
the scalar field in the different black hole spacetimes has been
investigated extensively.  It is found that for usual scalar field
the late-time evolution after the quasinormal oscillations  is
dominated by the form $t^{-(2l+3)}$ for the massless field
\cite{Price:1972,Hod:1998ja,Barack:2000} and by the oscillatory
inverse power-law form $t^{-(l+3/2)}\sin{\mu t}$ for the massive one
\cite{Koyama:2001,Hod:1998ma}. Moreover, the dynamical evolution of
the scalar field has also been considered in the cosmology, which
shows that the scalar field can be presented as the inflaton to
drive the inflation of the early Universe \cite{Guth:1981} and as
the dark energy to drive the accelerated expansion of the current
Universe \cite{Ratra:1988,Picon:1999,Caldwell:2002}. However, the
above investigations are limited to the case where the action has a
form
\begin{eqnarray}
S=\int d^4x \sqrt{-g}\bigg[\frac{R}{16\pi
G}+\frac{1}{2}\partial_{\mu}\psi\partial^{\mu}\psi+V(\psi)+\xi
R\psi^2\bigg]+S_m,\label{act1}
\end{eqnarray}
where $\psi$, $R$ and $V(\psi)$ are corresponding to scalar field,
Ricci scalar and scalar potential, respectively. In this action, the
coupling between the scalar field and the spacetime curvature
contains only the term $\xi R\psi^2$, which represents the coupling
between the scalar field and the Ricci scalar curvature.

Theoretically, the general form of the action with more couplings
between the scalar field and the spacetime curvature can be
expressed as
\begin{eqnarray}
S=\int d^4x \sqrt{-g}\bigg[f(\psi, R, R_{\mu\nu}R^{\mu\nu},
R_{\mu\nu\rho\sigma}R^{\mu\nu\rho\sigma}) +K(\psi,
\partial_{\mu}\psi\partial^{\mu}\psi,\nabla^2\psi, R^{\mu\nu}\partial_{\mu}\psi\partial_{\nu}\psi,\cdot\cdot\cdot)+V(\psi)\bigg]+S_m,\label{act2}
\end{eqnarray}
where $f$ and $K$ are arbitrary functions of the corresponding
variables. Obviously, the nonlinear functions $f$ and $K$ provide
the more non-minimal couplings between the scalar field and the
curvature of the background spacetime. These new couplings modify
the usual Klein-Gordon equation so that the motion equation for the
scalar field is no longer generally a second-order differential
equation in this case, which yields the more complicated behavior of
the scalar field in the background spacetime. By introducing the
derivative coupling term
$R_{\mu\nu}\partial^{\mu}\psi\partial^{\nu}\psi$, Amendola
\cite{Amendola:1993} studied recently the dynamical evolution of the
coupled scalar field in the cosmology and obtained some new
analytical inflationary solutions. Capozziello \textit{et al.}
\cite{Capozziello:2000,Capozziello:1999} investigated a more general
model with two coupling terms
$R\partial_{\mu}\psi\partial^{\nu}\psi$ and
$R_{\mu\nu}\partial^{\mu}\psi\partial^{\nu}\psi$, and found that the
de Sitter spacetime is an attractor solution in this case. Recently,
Sushkov found \cite{Sushkov:2009} that the equation of motion for
the scalar field can be reduced to second-order differential
equation when it is kinetically coupled to the Einstein tensor. This
means that the theory is a ``good" dynamical theory from the point
of view of physics. Moreover, Sushkov \cite{Sushkov:2009} also found
that in cosmology the problem of graceful exit from inflation  with
the derivative coupling term $
G^{\mu\nu}\partial_{\mu}\psi\partial_{\nu}$ has a natural solution
without any fine-tuned potential. Recently, Gao \cite{Gao:2010}
investigated the cosmic evolution of a scalar field with the kinetic
term coupling to more than one Einstein tensors, and found that the
scalar field behaves exactly as the pressureless matter if the
kinetic term is coupled to one Einstein tensor and acts nearly as a
dynamic cosmological constant if it couples with more than one
Einstein tensors. The similar investigations have been considered in
Refs.\cite{Granda:2009,Saridakis:2010}. We studied the greybody
factor and Hawking radiation for a scalar field  coupling to
Einstein's tensor in the background of Reissner-Nordstr\"{o}m black
hole spacetime and found that the presence of the coupling enhances
both the absorption probability and Hawking radiation of the black
hole \cite{Chen:2010}. These results may excite more efforts to be
focused on the study of the scalar field coupled with tensors in the
more general cases. The main purpose of this paper is to investigate
the dynamical evolution of the scalar perturbation coupling to the
Einstein tensor $G^{\mu\nu}$ in the Reissner-Nordstr\"{o}m black
hole spacetime and see the effect of the coupling on the stability
of the black hole.

The plan of our paper is organized as follows: in the following
section we will introduce the action of a scalar field coupling to
Einstein's tensor and derive its master equation in the
Reissner-Nordstr\"{o}m black hole spacetime. In Sec.III, we will
study the effect of the coupling on the quasinormal modes in the
weaker coupling, and then examine the stability of the black hole in
the stronger coupling. Finally, in the last section we will include
our conclusions.

\section{The wave equation of a scalar field coupling to Einstein's tensor in the Reissner-Nordstr\"{o}m black hole spacetime}

In order to study the dynamical evolution of a scalar field coupling
to Einstein's tensor in a black hole spacetime, we must first obtain
its wave equation in the background. The action of the scalar field
coupling to the Einstein's tensor $G^{\mu\nu}$ in the curved
spacetime has a form \cite{Sushkov:2009},
\begin{eqnarray}
S=\int d^4x \sqrt{-g}\bigg[\frac{R}{16\pi
G}+\frac{1}{2}\partial_{\mu}\psi\partial^{\mu}\psi+\frac{\eta}{2}G^{\mu\nu}\partial_{\mu}\psi\partial_{\nu}\psi\bigg].\label{acts}
\end{eqnarray}
The coupling between Einstein's tensor $G^{\mu\nu}$ and the scalar
field $\psi$ is represented by the term
$\frac{\eta}{2}G^{\mu\nu}\partial_{\mu}\psi\partial_{\nu}\psi$,
where $\eta$ is coupling constant with dimensions of length-squared.

Varying the action (\ref{acts}) with respect to $\psi$, one can find
the wave equation of a scalar field coupling to Einstein's tensor
can be expressed as \cite{Sushkov:2009,Chen:2010}
\begin{eqnarray}
\frac{1}{\sqrt{-g}}\partial_{\mu}\bigg[\sqrt{-g}\bigg(g^{\mu\nu}+\eta
G^{\mu\nu}\bigg)\partial_{\nu}\psi\bigg] =0.\label{WE}
\end{eqnarray}
Obviously, the dynamical evolution of a scalar field depends on the
the Einstein's tensor $G^{\mu\nu}$ and the coupling constant $\eta$.
Since all the components of the tensor $G^{\mu\nu}$ vanish in the
Schwarzschild black hole spacetime, we cannot probe the effect of
the coupling term on the dynamical behavior of the scalar
perturbation. In the general relative theory, the simplest black
hole with the non-zero components of the tensor $G^{\mu\nu}$ is
Reissner-Nordstr\"{o}m one, whose metric has a form
\begin{eqnarray}
ds^2&=-&fdt^2+\frac{1}{f}dr^2+r^2
d\theta^2+r^2\sin^2{\theta}d\phi^2,\label{m1}
\end{eqnarray}
with
\begin{eqnarray}
f=1-\frac{2M}{r}+\frac{q^2}{r^2},
\end{eqnarray}
where $M$ is the mass and $q$ is the charge of the black hole. The
Einstein's tensor $G^{\mu\nu}$ for the metric (\ref{m1}) has a form
\begin{eqnarray}
G^{\mu\nu}= \frac{q^2}{r^4} \left(\begin{array}{cccc}
 -\frac{1}{f}&&&\\
 &f&&\\
 &&-\frac{1}{r^2}&\\
 &&&-\frac{1}{r^2\sin^2\theta}
\end{array}\right).
\end{eqnarray}
Defining tortoise coordinate $dr_*=1/f(r)dr$ and separating
$\psi(t,r,\theta,\phi)=\frac{e^{-i\omega
t}R(r)Y_{lm}(\theta,\phi)r}{\sqrt{r^4+\eta q^2}}$,  we can obtain
the radial equation for the scalar perturbation coupling to
Einstein's tensor in the Reissner-Nordstr\"{o}m black hole spacetime
\begin{eqnarray}
\frac{d^2R(r)}{dr^2_*}+[\omega^2-V(r)]R(r)=0,\label{radial}
\end{eqnarray}
with the effective potential
\begin{eqnarray}
V(r)=f\bigg(\frac{r^4-\eta q^2}{r^4+\eta
q^2}\bigg)\bigg[\frac{l(l+1)}{r^2}+\frac{f'}{r}\bigg]+\frac{f^2}{r^2}\frac{2\eta
q^2(3r^4+\eta q^2)}{(r^4+\eta q^2)^2}.\label{ev}
\end{eqnarray}
Obviously, as the coupling constant $\eta=0$ the radial equation
(\ref{radial}) reduces to that of the scalar one without coupling to
Einstein's tensor. In the case $\eta\neq0$, the coupling constant
$\eta$ emerges in the effective potential, which means that coupling
between the scalar perturbation and Einstein's tensor will change
the dynamical evolution of the scalar perturbation in the background
spacetime.
\begin{figure}[ht]
\begin{center}
\includegraphics[width=5.5cm]{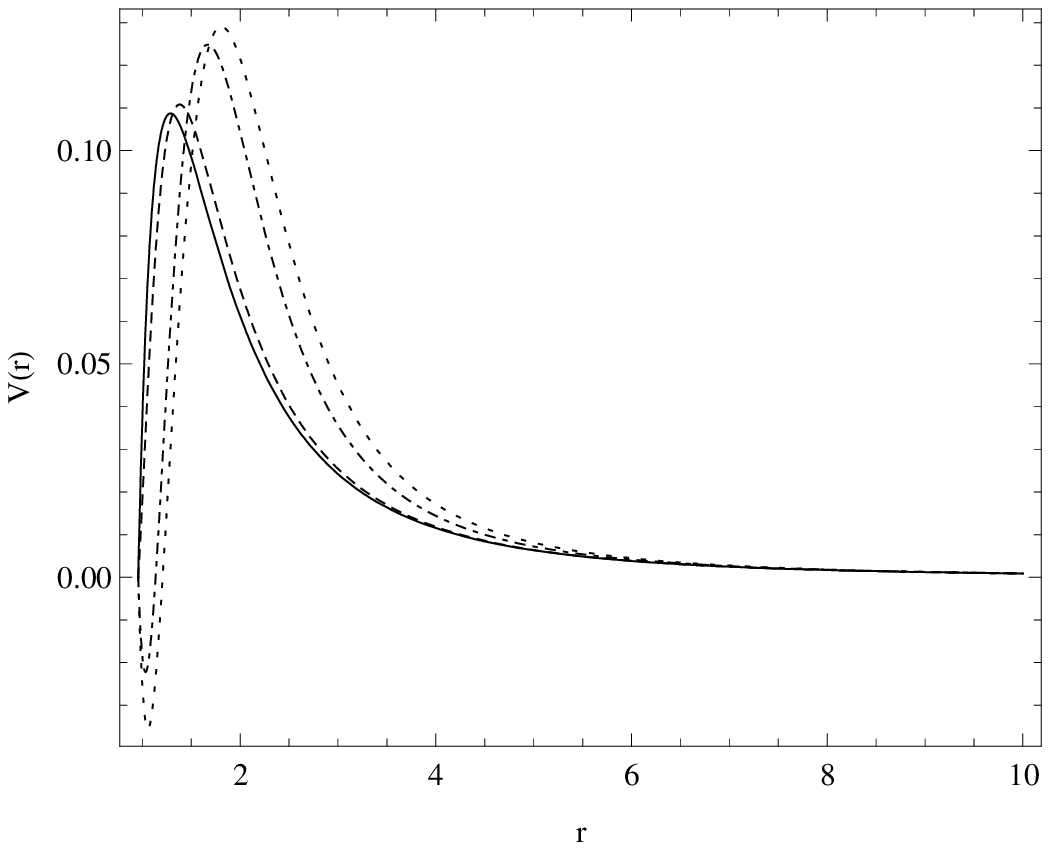}\includegraphics[width=5.5cm]{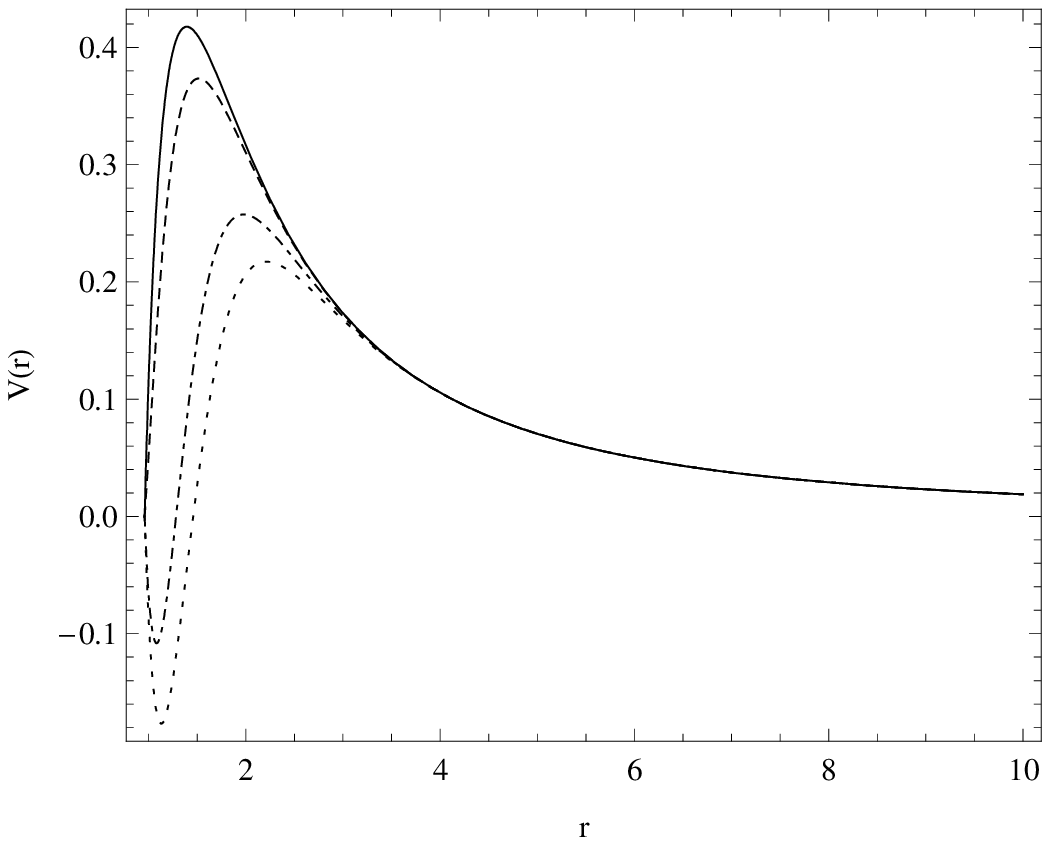}\includegraphics[width=5.5cm]{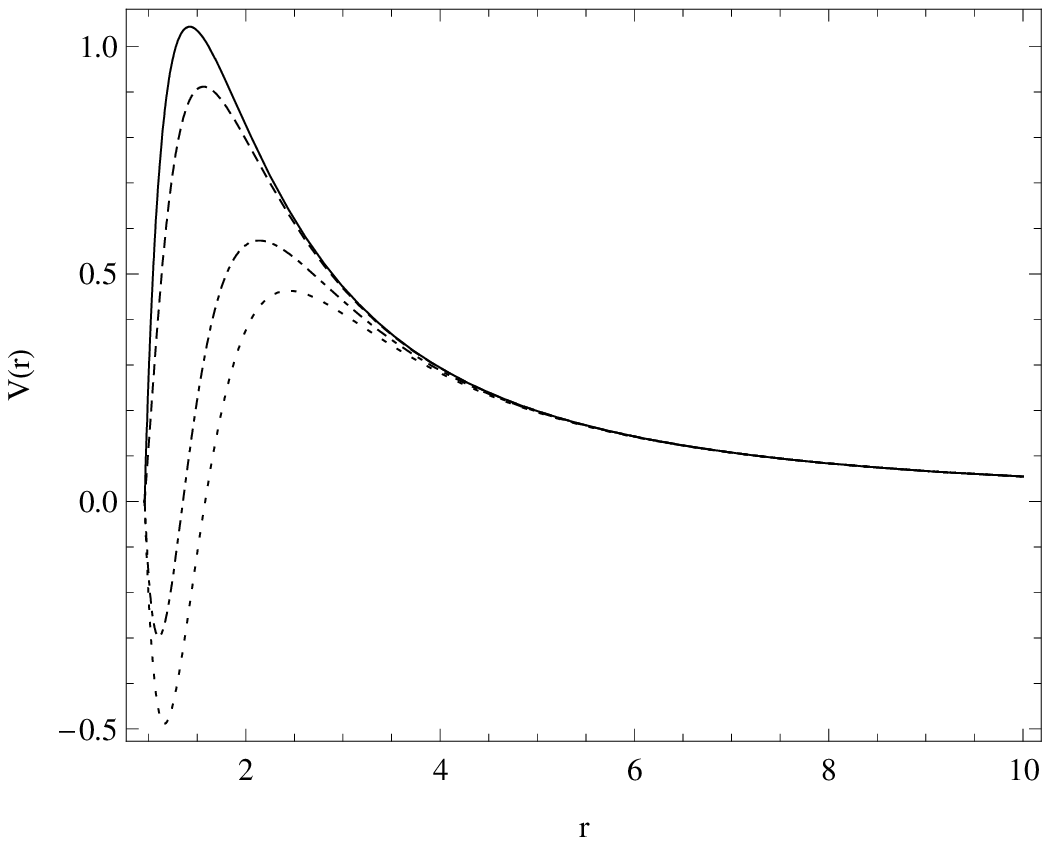}
\caption{Variety of the effective potential $V(r)$ with the polar
coordinate $r$ for fixed $l=0$ (left), $l=1$ (middle) and $l=2$
(right). The solid, dashed, dash-dotted and dotted lines are
corresponding to the cases with $\eta=0,\;10,\;100,\;200$,
respectively. We set $2M=1$ and $q=0.2$.}
\end{center}
\label{fig1}
\end{figure}
In Fig.1, we plot the changes of the effective potential $V(r)$ with
the coupling constant $\eta$ for fixed $l$ and $q$. With increase of
$\eta$, the peak height of the potential barrier increases for $l=0$
and decreases for other values of $l$. Moreover, one can find that
for the smaller $\eta$ the effective potential $V(r)$ is positive
definite everywhere outside the black hole event horizon. This
implies that the solution of the wave equation (\ref{WE}) is bounded
and the black hole is stable in this case. However, for the larger
$\eta$, we find that the effective potential $V(r)$ has negative
gap, and then the stability is not guaranteed. In the following
section, we will check those values of $\eta$ for which the negative
gap is present and study the stability of the black hole when the
scalar perturbation is coupling to Einstein's tensor.

\section{The instability of scalar field coupling to Einstein's
tensor in the background of a Reissner-Nordstr\"{o}m black hole}

In this section, we first consider the quasinormal modes in the
weaker coupling case in which the effective potential $V(r)$ is
positive definite and study the effects of the coupling on the
quasinormal frequencies. Then, we shall study the evolution of the
scalar field coupling to Einstein's tensor in time domain using a
numerical characteristic integration method \cite{Gundlach:1994} and
check the instability of the black hole in the stronger coupling.

Let us now to study the effects of the coupling constant on the
massless scalar quasinormal modes in the Reissner-Nordstr\"{o}m
black hole spacetime in the weaker coupling case. In Fig.2 and 3, we
present the fundamental quasinormal modes ($n=0$) evaluated by the
third-order WKB approximation method \cite{Schutz:1985,Iyer:1987}.
It is shown that with the increase of the coupling constant $\eta$
the real parts of the quasinormal frequencies decrease and the
absolute values of imaginary parts increases for fixed $l$ and $q$.
\begin{figure}[ht]
\begin{center}
\includegraphics[width=5cm]{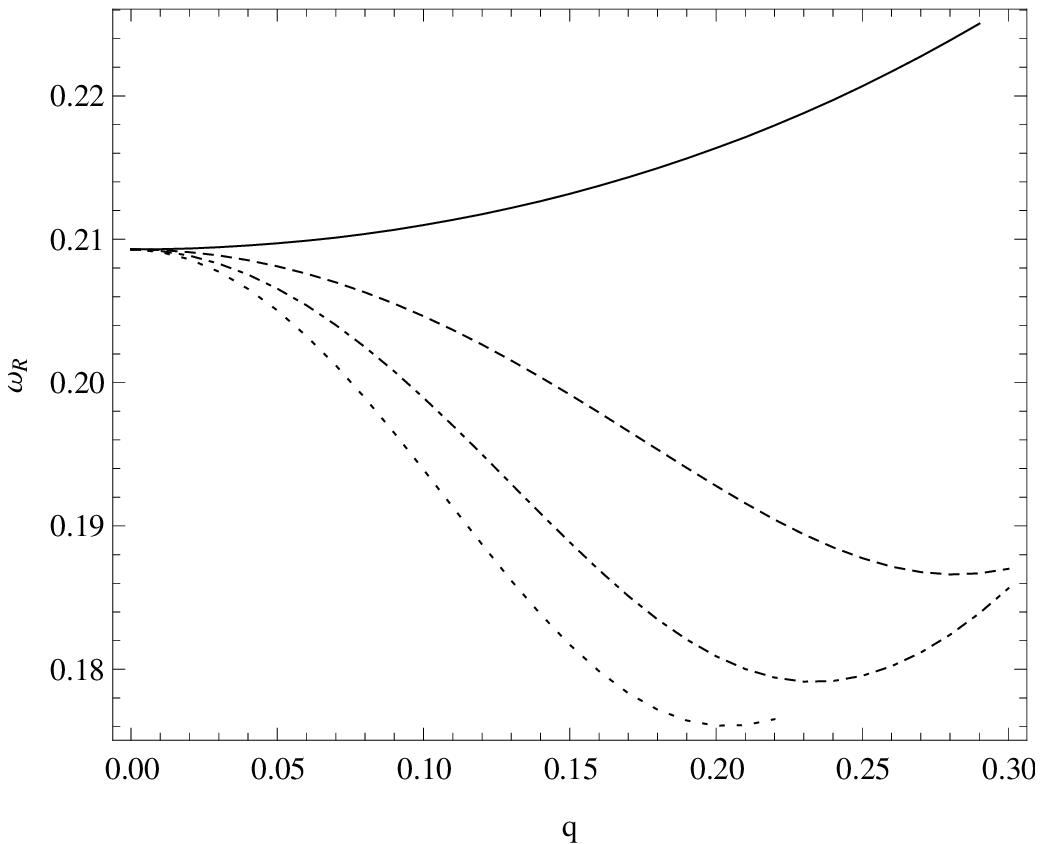}\includegraphics[width=5cm]{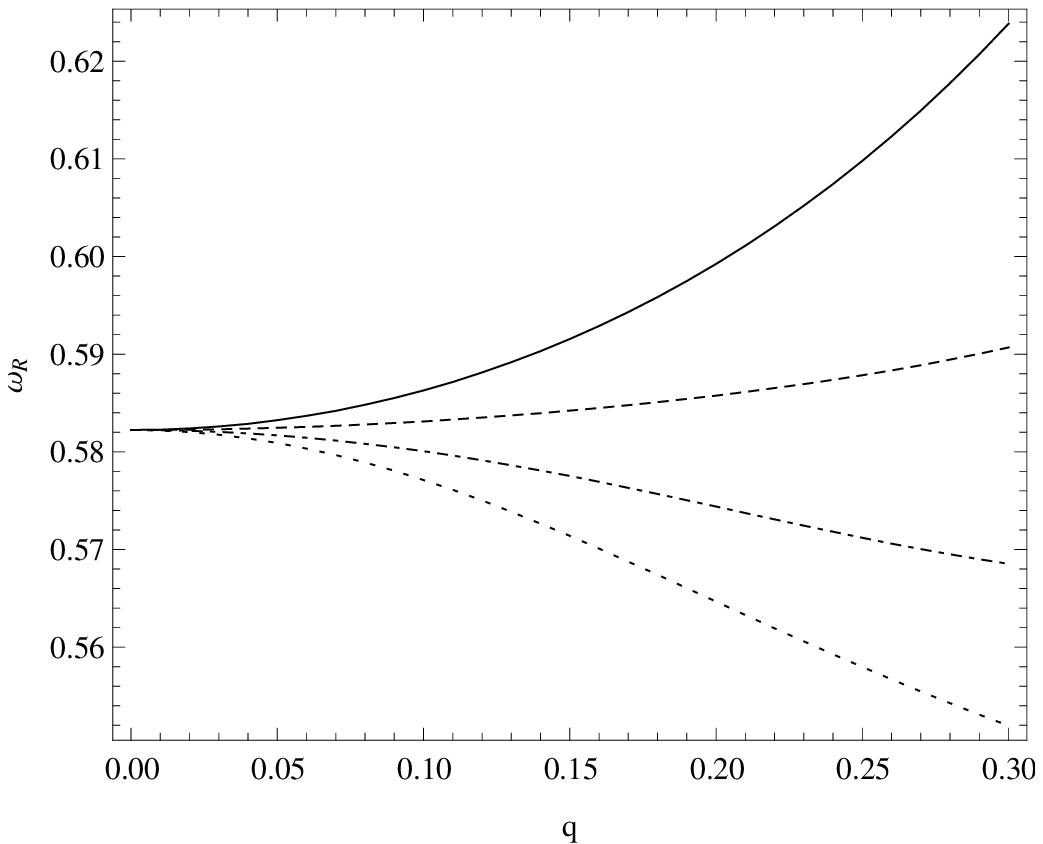}\includegraphics[width=5cm]{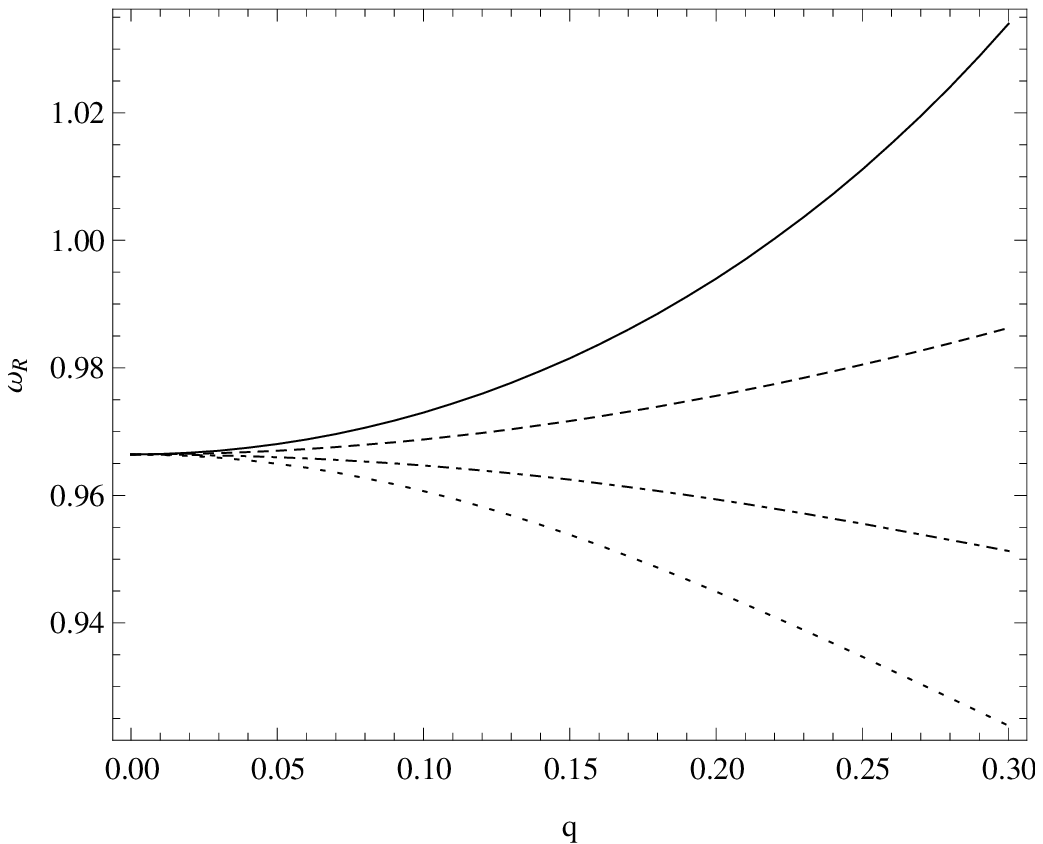}
\caption{Variety of the real parts of the fundamental quasinormal
modes with $q$ for scalar field coupling to Einstein's tensor in the
Reissner-Nordstr\"{o}m black hole spacetime. The figures from left
to right are corresponding to $l=0,\;1$ and $2$. The solid, dashed,
dash-dotted and dotted lines are corresponding to the cases with
$\eta=0,\;2,\;4,\;6$, respectively. We set $2M=1$.}
\end{center}
\label{fig2}
\end{figure}
\begin{figure}[ht]
\begin{center}
\includegraphics[width=5cm]{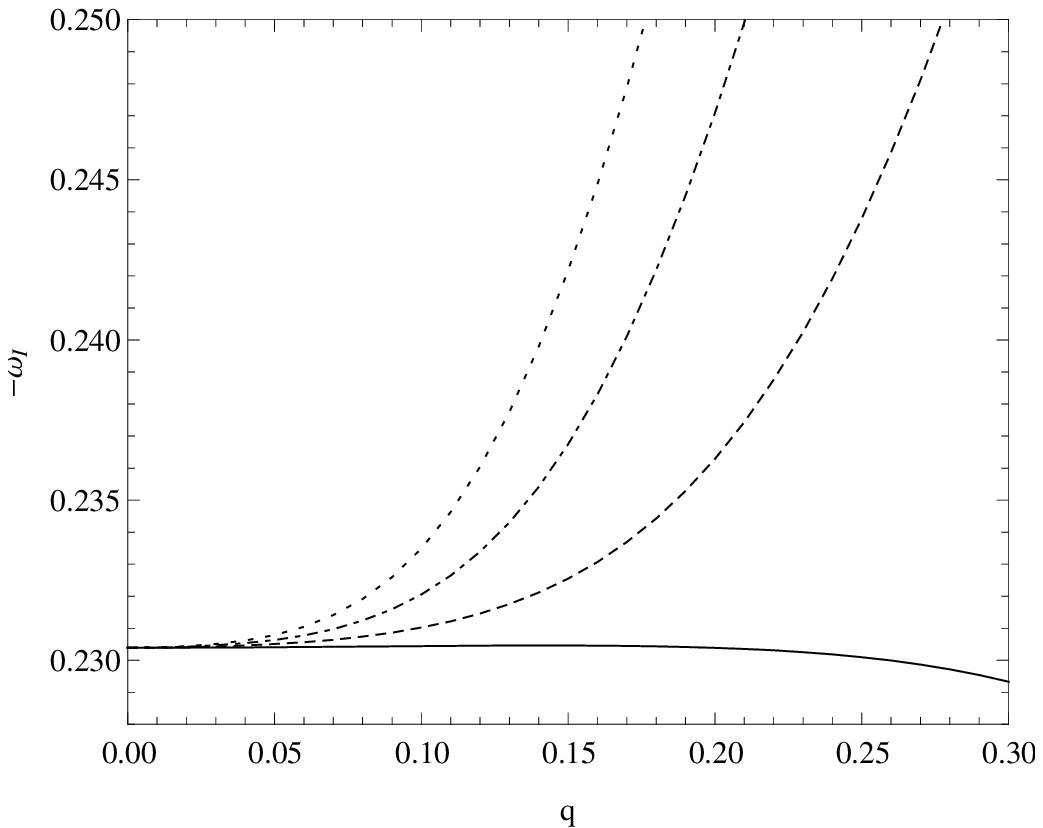}\includegraphics[width=5cm]{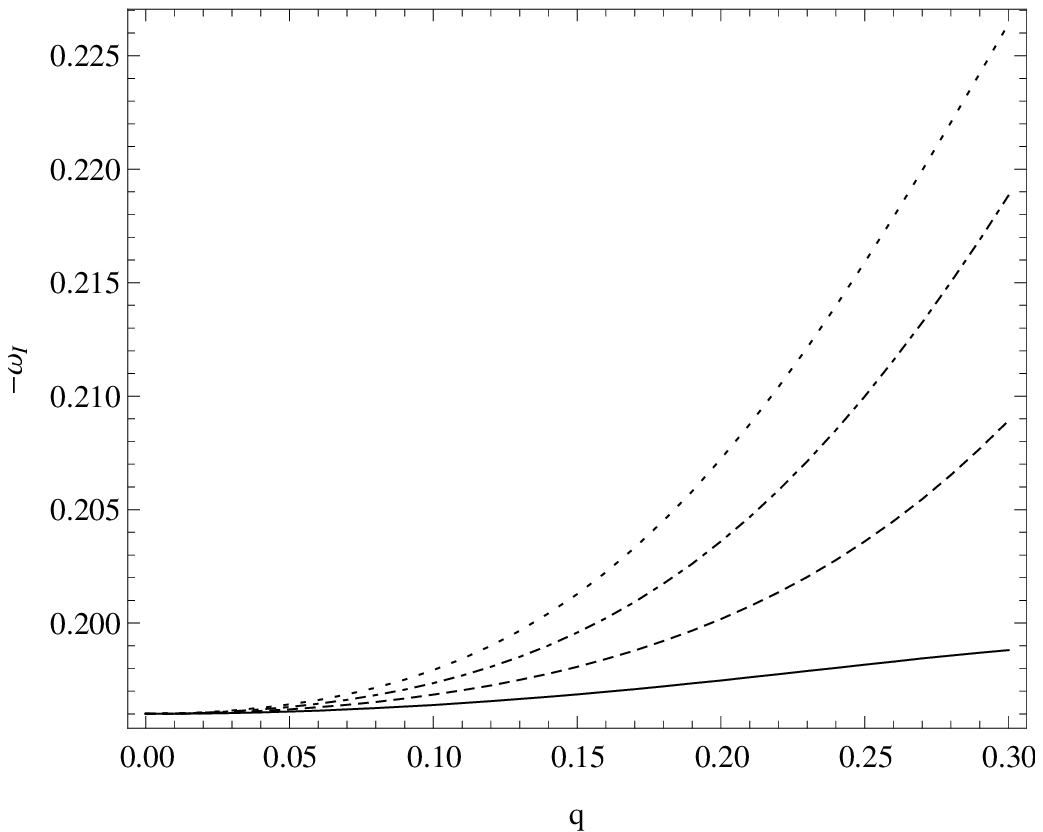}\includegraphics[width=5cm]{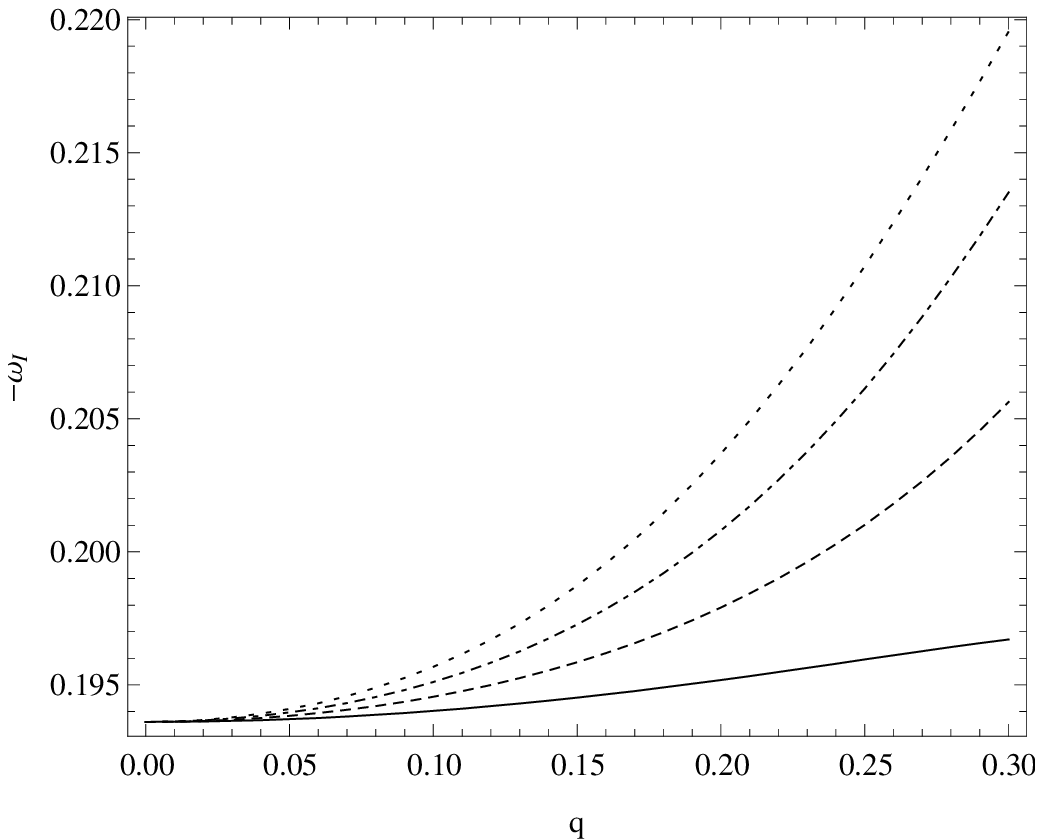}
\caption{Variety of the absolute value of imaginary parts of the
fundamental quasinormal modes with $q$ for scalar field coupling to
Einstein's tensor in the Reissner-Nordstr\"{o}m black hole
spacetime. The figures from left to right are corresponding to
$l=0,\;1$ and $2$. The solid, dashed, dash-dotted and dotted lines
are corresponding to the cases with $\eta=0,\;2,\;4,\;6$,
respectively. We set $2M=1$.}
\end{center}
\label{fig3}
\end{figure}
This means that the presence of the coupling parameter $\eta$ makes
the decay of the scalar perturbation more quickly in this case. From
Fig.2 and 3, one can easily obtain that with increase of $q$ the
real parts $\omega_R$ increases for the smaller $\eta$ and decreases
for the larger $\eta$. The changes of the absolute values of
imaginary parts with $q$ are more complicated. For $l=0$, it
decreases for the smaller $\eta$ and increases for the larger
$\eta$, while for other values of $l$, it increase with $q$ for all
$\eta$. These results imply that the presence of the coupling terms
modifies the standard results in the quasinormal modes of the scalar
perturbations in the background of a Reissner-Nordstr\"{o}m black
hole.

We are now in a position to study the dynamical evolution of the
scalar field coupling to Einstein's tensor in time domain and
examine the stability of the black hole in the stronger coupling
cases. Adopting to the light-cone variables $u=t-r_*$ and $v=t+r_*$,
one can find that the wave equation
\begin{eqnarray}
-\frac{\partial^2\psi}{\partial t^2}+\frac{\partial^2\psi}{\partial
r_*^2}=V(r)\psi,
\end{eqnarray}
can be rewritten as
 \begin{eqnarray}
4\frac{\partial^2\psi}{\partial u\partial
v}+V(r)\psi=0.\label{wbes1}
\end{eqnarray}
\begin{figure}[ht]\label{fig8}
\begin{center}
\includegraphics[width=5.5cm]{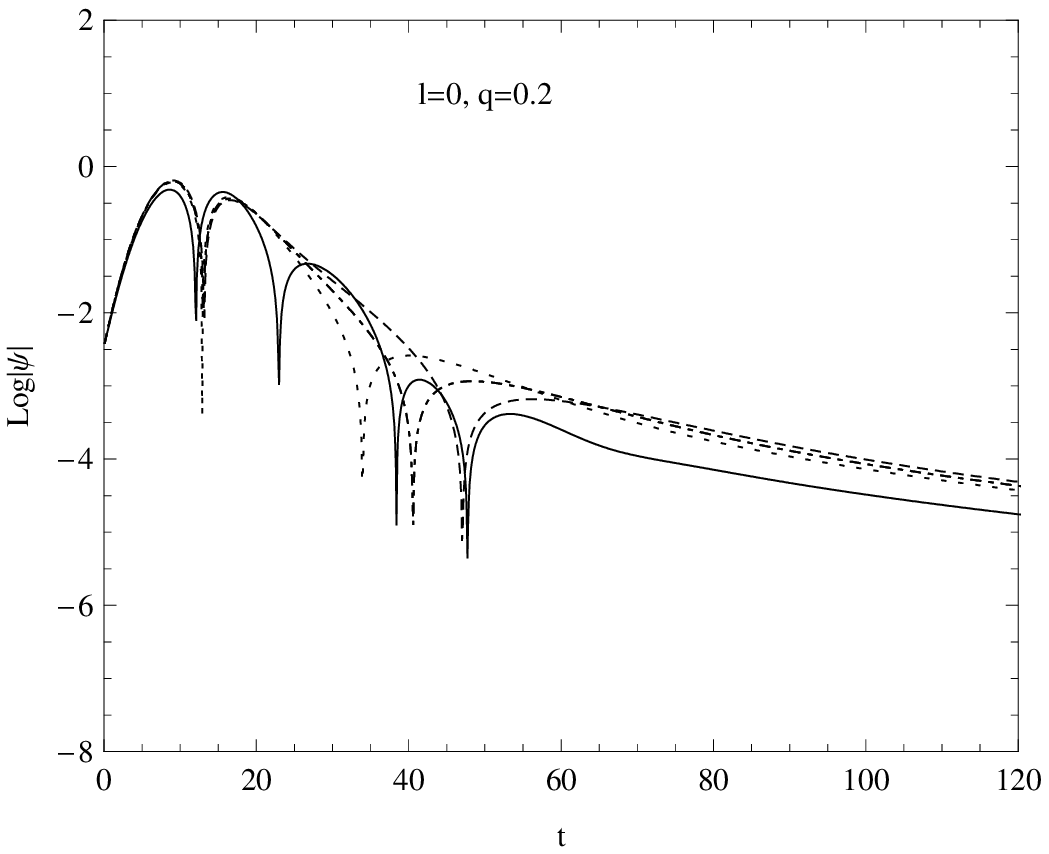}\;\;\includegraphics[width=5.5cm]{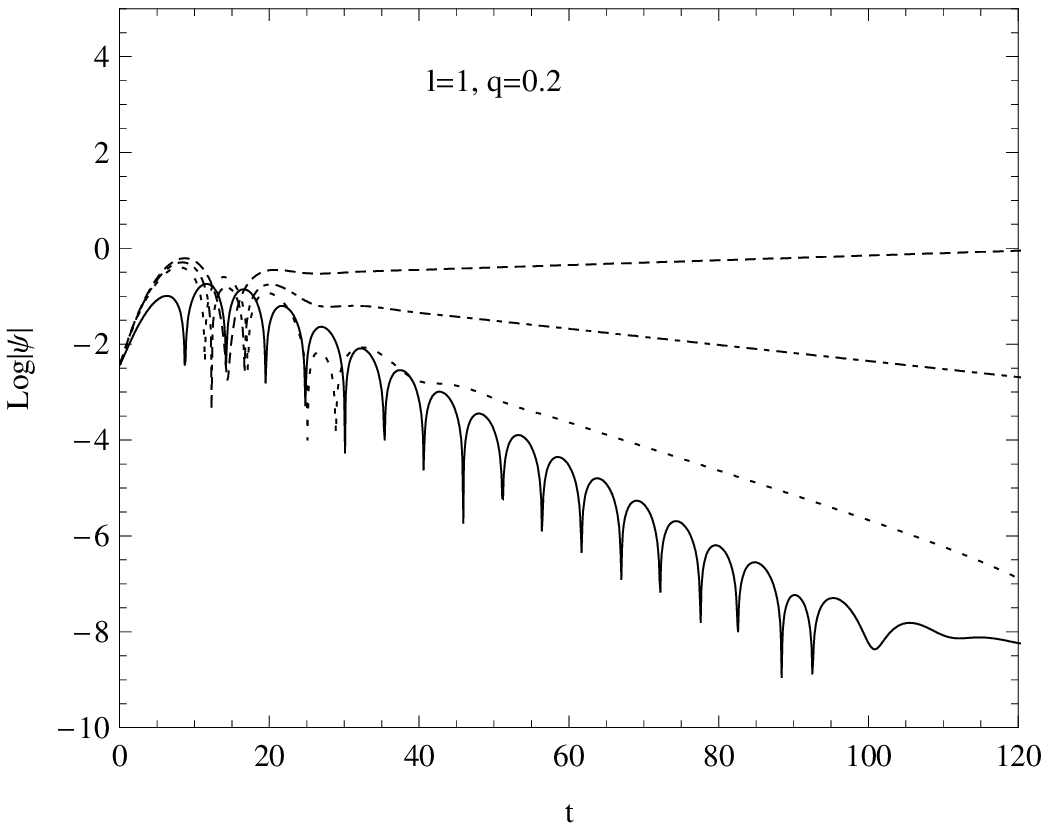}\;\;\includegraphics[width=5.5cm]{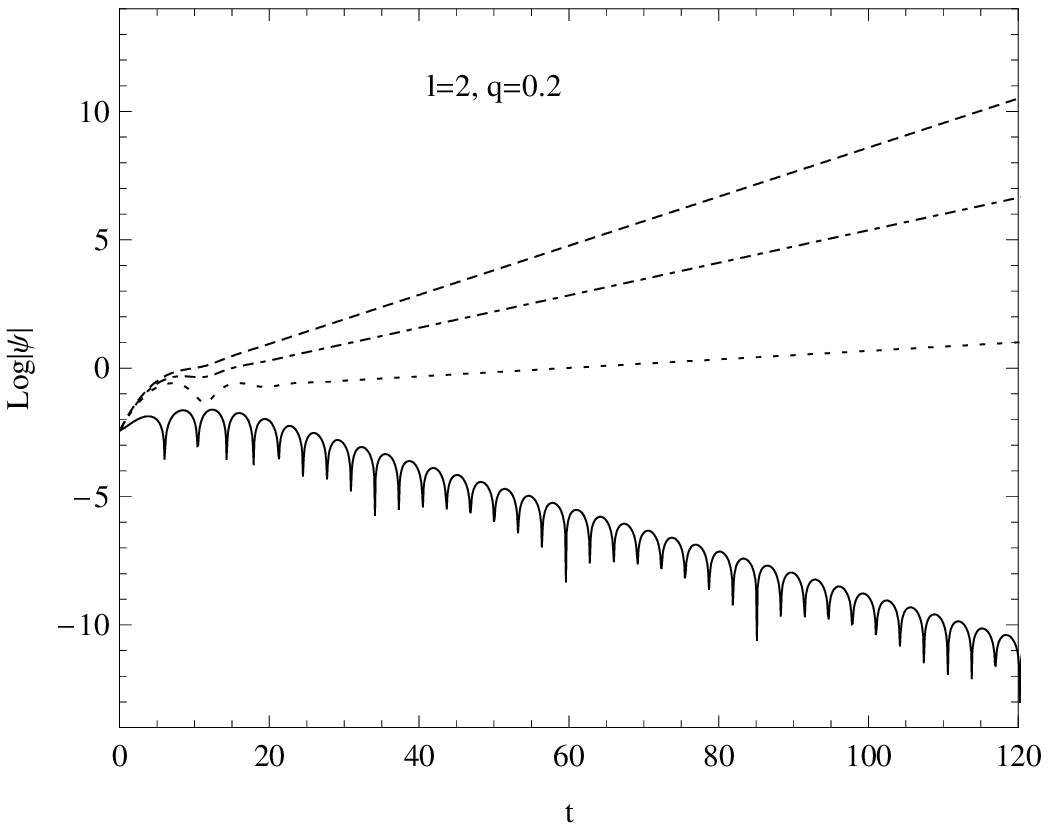}
\caption{The dynamical evolution of a scalar field coupling to
Einstein's tensor in the background of a Reissner-Nordstr\"{o}m
black hole spacetime. The figures from left to right are
corresponding to $l=0,\;1$ and $2$. The solid, dashed, dash-dotted
and dotted lines are corresponding to the cases with
$\eta=1,\;60,\;80,\;100$, respectively. We set $2M=1$. The constants
in the Gauss pulse (\ref{gauss}) $v_c=10$ and $\sigma=3$.}
\end{center}
\end{figure}
This two-dimensional wave equation (\ref{wbes1}) can be integrated
numerically by using the finite difference method suggested in
\cite{Gundlach:1994}. In terms of Taylor's theorem, it can be
discretized as
\begin{eqnarray}
\psi_N=\psi_E+\psi_W-\psi_S-\delta u\delta v
V(\frac{v_N+v_W-u_N-u_E}{4})\frac{\psi_W+\psi_E}{8}+O(\epsilon^4)=0,\label{wbes2}
\end{eqnarray}
where we have used the following definitions for the points: $N$:
$(u+\delta u, v+\delta v)$, $W$: $(u + \delta u, v)$, $E$: $(u, v +
\delta v)$ and $S$: $(u, v)$. The parameter $\epsilon$ is an overall
grid scalar factor, so that $\delta u\sim\delta v\sim\epsilon$.
Since the behavior of the wave function is not sensitive to the
choice of initial data, we can set $\psi(u, v=v_0)=0$ and use a
Gaussian pulse as an initial perturbation, centered on $v_c$ and
with width $\sigma$ on $u=u_0$ as
\begin{eqnarray}
\psi(u=u_0,v)=e^{-\frac{(v-v_c)^2}{2\sigma^2}}.\label{gauss}
\end{eqnarray}
\begin{figure}[ht]\label{fig9}
\begin{center}
\includegraphics[width=6cm]{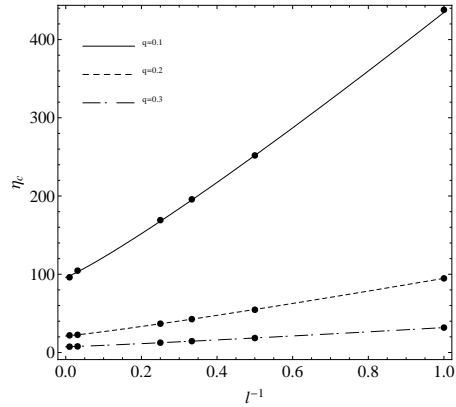}
\caption{The change of the threshold value $\eta_c$ with the inverse
multipole number $l^{-1}$ for fixed $q$. The points
$l=1,2,3,4,32,100$ were fitted by the function
$\eta_c=al^{-1.12}+b$. The values of $(a,b)$ for $q=0.1,\;0.2$ and
$0.3$ are $(338.71,96.00),\; (73.72, 21.10)$ and $(24.52,7.31)$,
respectively.}
\end{center}
\end{figure}
In fig.4, we present the dynamical evolution of the scalar field
coupling to Einstein's tensor in the background of a
Reissner-Nordstr\"{o}m black hole. For the coupling constant
$\eta=1$, the decay of the coupling scalar field is similar to that
of the scalar one without coupling to Einstein's tensor, which
indicates that the black hole is stable in the weaker coupling. It
is expectable because the effective potential $V(r)$ is positive
definite in this cases. For $l=0$, we also note that the scalar
field always decays for any value of the coupling constant $\eta$.
This means that the lowest $l$ are stable, which can be explained by
a fact that for $l=0$, the higher $\eta$ raise up the peak of the
potential barrier so that the potential is always positive definite.
Moreover, for the higher multipole numbers $l$, we find that the
scalar field grows with exponential rate as the coupling constant
$\eta$ is larger than the critical value $\eta_c$, which means that
the instability occurs in this case. The main reason is that for
$l\neq0$ the large $\eta$ drops down the peak of the potential
barrier and increases the negative gap near the black hole horizon
so that the potential could be non-positive definite. In the
instability region, the larger $\eta$, the instability growth occurs
at the earlier times, and the growth rate is the stronger.
Furthermore, we plotted the change of the threshold value $\eta_c$
with $l$ in fig.5, and found that the threshold value can be fitted
best by the function
\begin{eqnarray}
\eta_c\simeq al^{-1.12}+b,\label{ncs}
\end{eqnarray}
\begin{figure}[ht]\label{fig10}
\begin{center}
\includegraphics[width=6cm]{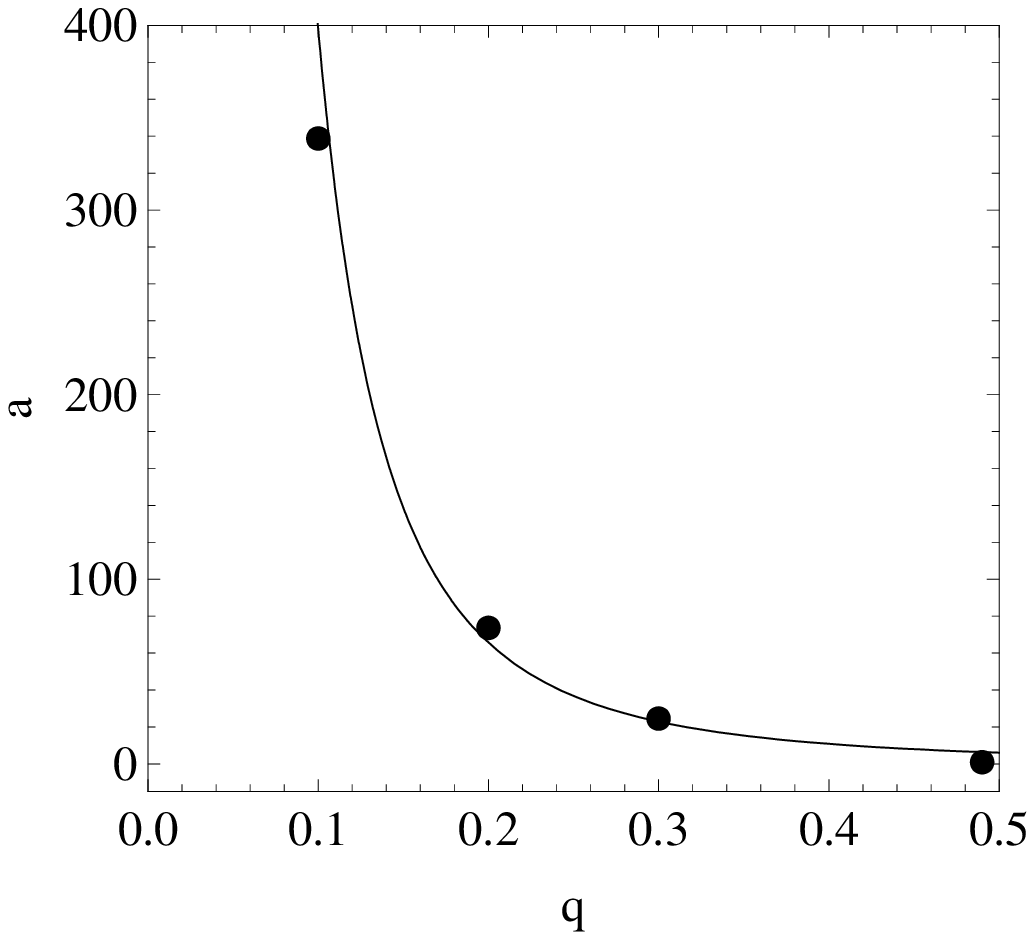}\;\;\;\;\;\;\includegraphics[width=6cm]{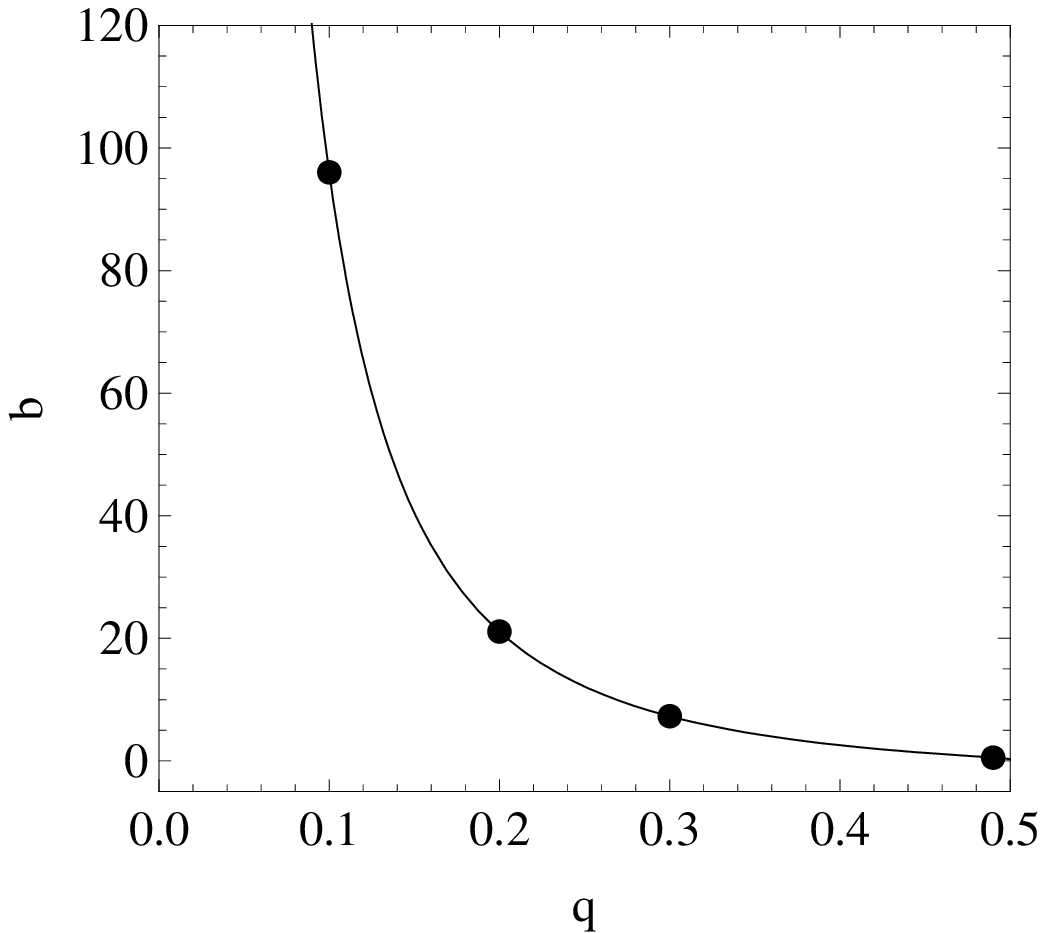}
\caption{The changes of the numerical constants $a, b$ with the
charge $q$, which were fitted by the functions $a= q^{-2.6}$ and $b=
r^4_+/q^2$, respectively.}
\end{center}
\end{figure}
where $a$ and $b$ are numerical constants. It is easy to obtain that
the lowest $l$ are stable because the threshold value
$\eta_c\rightarrow\infty$ as $l\rightarrow0$. Moreover, for the
higher $l$ and smaller $q$, we have the smaller threshold value
$\eta_c$ at which instability happens. The varieties of the
numerical constants $a, b$ with $q$ are presented in fig.6, which
shows that the values of $a, b$ are fitted best by the functions
$a\simeq q^{-2.6}$ and $b\simeq r^4_+/q^2$, respectively. Thus, as
the charge $q$ vanishes, the threshold value $\eta_c$ tends to
infinite for arbitrary $l$, which means that the Schwarzschild black
hole is stable when it is perturbed by a scalar field coupling to
Einstein's tensor. Actually, since all the components of the
Einstein's tensor disappear in the Schwarzschild black hole
spacetime, the dynamical evolution of the coupling scalar field is
consistent with that of the scalar one without coupling to
Einstein's tensor. For the extreme black hole, we find from fig.6
and Eq.(\ref{ncs}) that the threshold value $\eta_c$ is the minimum
for arbitrary $l\neq 0$, which implies that the instability happens
more easily in the extreme black hole. From the previous
discussions, we can obtain that in the limit $l\rightarrow\infty$
the threshold value $\eta_c\rightarrow b$ and the effective
potential (\ref{ev}) has the form
\begin{eqnarray}
V(r)|_{l\rightarrow\infty}=f\bigg(\frac{r^4-\eta q^2}{r^4+\eta
q^2}\bigg)\frac{l(l+1)}{r^2}.\label{evl}
\end{eqnarray}
According to the method suggested in \cite{Gleiser:2005}, the
integration
\begin{eqnarray}
\int^{\infty}_{r_+}\frac{V(r)|_{l\rightarrow\infty}}{f}dr=\int^{\infty}_{r_+}\bigg(\frac{r^4-\eta
q^2}{r^4+\eta q^2}\bigg)\frac{l(l+1)}{r^2}dr,
\end{eqnarray}
is positive definite as $\eta<r^4_+/q^2$. It implies that the
threshold value has a form $\eta_c=r^4_+/q^2$ as
$l\rightarrow\infty$, which is consistent with the form of the
numerical constant $b$ obtained in the previous numerical
calculation.

\section{summary}
In this paper, we have studied the dynamical evolution of a scalar
field coupling to Einstein's tensor in the background of
Reissner-Nordstr\"{o}m black hole. Our results show that the the
coupling constant $\eta$ imprints in the wave dynamics of a scalar
perturbation. For the multipole number $l=0$, we find that the
scalar field always decays for arbitrary coupling constant $\eta$.
For $l\neq0$, the instability occurs when $\eta$ is larger than the
critical value $\eta_c$. Moreover, for the higher $l$, we have the
smaller threshold value $\eta_c$. In the weak coupling (i.e.,
$\eta\ll\eta_c$) case, we find that with the increase of the
coupling constant $\eta$ the real part of the fundamental
quasinormal frequencies decreases and the absolute value of
imaginary parts increases for fixed $l$ and $q$. With increase of
$q$ the real part $\omega_R$ increases for the smaller $\eta$ and
decreases for the larger $\eta$.  For $l=0$, the absolute value of
imaginary parts decreases for the smaller $\eta$ and increases for
the larger $\eta$, while for other values of $l$, it increases with
$q$ for all $\eta$. Moreover, we find that the threshold value can
be fitted best by the function $\eta_c\simeq al^{-1.12}+b$ and the
numerical constants $a, b$ decrease with the charge $q$. These rich
dynamical properties of the scalar field coupling to Einstein's
tensor, which could provide a way to detect whether there exist a
coupling between the scalar field and Einstein's tensor or not. It
would be of interest to generalize our study to other black hole
spacetimes, such as rotating black holes etc. Work in this direction
will be reported in the future.

\section{\bf Acknowledgments}

This work was  partially supported by the National Natural Science
Foundation of China under Grant No.10875041,  the Program for
Changjiang Scholars and Innovative Research Team in University
(PCSIRT, No. IRT0964) and the construct program of key disciplines
in Hunan Province. J. Jing's work was partially supported by the
National Natural Science Foundation of China under Grant Nos.
10875040 and 10935013; 973 Program Grant No. 2010CB833004 and the
Hunan Provincial Natural Science Foundation of China under Grant
No.08JJ3010.


\vspace*{0.2cm}
\begin{thebibliography}{99}
\baselineskip=0.6 cm

\bibitem{Chandrasekhar:1975}S. Chandrasekhar and S. Detweller, Proc. R. Soc. Lond. A {\bf344}, 441
(1975).
\bibitem{Nollert:1999} H. P. Nollert, Class. Quantum Grav. {\bf16}, R159 (1999).

\bibitem{Kokkotas:1999} K. D. Kokkotas and B. G. Schmidt, Living Rev. Rel. {\bf2}, 2 (1999).

\bibitem{Hod:1998} S. Hod, Phys. Rev. Lett. {\bf81}, 4293 (1998).
\bibitem{Dreyer:2003} O. Dreyer, Phys. Rev. Lett. {\bf90}, 081301 (2003).

\bibitem{Corichi:2003} A. Corichi, Phys. Rev. D {\bf67},  087502 (2003);
 L. Motl, Adv. Theor. Math. Phys. {\bf6},  1135 (2003);
L. Motl and A. Neitzke, Adv. Theor. Math. Phys. {\bf7}, 307 (2003);
A. Maassen van den Brink, J. Math. Phys. {\bf45}, 327 (2004); G.
Kunstatter, Phys. Rev. Lett. {\bf 90}, 161301 (2003); N. Andersson
and C. J. Howls, Class. Quantum Grav. {\bf21}, 1623 (2004); V.
Cardoso, J. Natario and R. Schiappa, J. Math. Phys. {\bf45}, 4698
(2004); J. Natario and R. Schiappa, Adv. Theor. Math. Phys. {\bf8},
1001 (2004); V. Cardoso and J. P. S. Lemos, Phys. Rev. D {\bf67},
084020 (2003).


\bibitem{Maldacena:1998} J. Maldacena, Adv. Theor. Math. Phys. {\bf2}, 231 (1998).

\bibitem{Witten:1998} E. Witten, Adv. Theor. Math. Phys. {\bf2}, 253 (1998).

\bibitem{Horowitz:2000} G. T. Horowitz and V. E. Hubeny, Phys. Rev. D {\bf62}, 024027 (2000);
B. Wang, C. Y. Lin and E. Abdalla, Phys. Lett. B {\bf 481}, 79
(2000) ; J. M. Zhu, B. Wang and E. Abdalla, Phys. Rev. D {\bf63},
124004 (2001); V. Cardoso and J. P. S. Lemos, Phys. Rev. D {\bf63},
124015 (2001); V. Cardoso and J. P. S. Lemos, Phys. Rev. D {\bf64},
084017 (2001); E. Berti and K. D. Kokkotas, Phys. Rev. D {\bf67},
064020 (2003); E. Winstanley, Phys. Rev. D {\bf64}, 104010 (2001);
J. S. F. Chan and R. B. Mann, Phys. Rev. D {\bf59}, 064025 (1999).

\bibitem{Gregory:1993} R. Gregory and R. Laflamme, Phys. Rev. Lett. {\bf 70}, 2837 (1993); R. Gregory and R. Laflamme, Nucl. Phys. B {\bf 428}, 399
(1994).
\bibitem{Harmark:2007} T. Harmark, V. Niarchos and N. A. Obers, Class. Quant. Grav. {\bf 24}, R1
(2007).
\bibitem{Konoplya:2008} R. A. Konoplya, K. Murata, Jiro Soda and A. Zhidenko, Phys. Rev. D {\bf 78}, 084012
(2008); J. L. Hovdebo and R. C. Myers,  Phys. Rev. D {\bf 73},
084013 (2006).
\bibitem{Chen:2009} S. B. Chen and J. L. Jing, JHEP {\bf03}, 081 (2009).

\bibitem{Price:1972} R. H. Price, Phys. Rev. D {\bf5}, 2419 (1972).
\bibitem{Hod:1998ja} S. Hod and T. Piran, Phys. Rev. D {\bf58}, 024017 (1998).
\bibitem{Barack:2000} L. Barack, Phys. Rev. D {\bf61}, 024026 (2000); L. M. Burko and G.
Khanna, Phys. Rev. D {\bf67}, 081502 (2003); E. S. C. Ching, P. T.
Leung, W. M. Suen and K. Young, Phys. Rev. D {\bf52}, 2118 (1995).


\bibitem{Koyama:2001}H. Koyama and A. Tomimatsu,
Phys. Rev. D {\bf63},  064032 (2001); H. Koyama and A. Tomimatsu,
Phys. Rev. D {\bf64},  044014 (2001); R. Moderski and M. Rogatko,
Phys. Rev. D {\bf64},  044024 (2001); R. Moderski and M. Rogatko,
Phys. Rev. D {\bf63}, 084014 (2001); R. Moderski and M. Rogatko,
Phys. Rev. D {\bf72}, 044027 (2005); S. Hod and T. Piran, Phys. Rev.
D {\bf58},  044018 (1998). S. B. Chen and J. L. Jing, Mod. Phys.
Lett. A {\bf23},  35 (2008); S. B. Chen, B. Wang and R. K. Su, Int.
J. Mod. Phys. A {\bf16}, 2502 (2008).

\bibitem{Hod:1998ma} S. Hod, Phys. Rev. D {\bf58}, 104022 (1998) ;
 L. Barack and A. Ori, Phys. Rev. Lett. {\bf82}, 4388 (1999) ;
 W. krivan, Phys. Rev. D {\bf60}, 101501(R)  (1999);
 Q. Y. Pan and J. L. Jing, Chin. Phys. Lett. {\bf21}, 1873 (2004).


\bibitem{Guth:1981} A. H. Guth, Phys. Rev. D {\bf23}, 347 (1981).
1a
\bibitem{Ratra:1988} B. Ratra and J. Peebles, Phys. Rev. D {\bf37}, 3406 (1988);
C. Wetterich,  Nucl. Phys. B {\bf 302}, 668 (1988); R. R. Caldwell,
R. Dave and P. J. Steinhardt, Phys. Rev. Lett. {\bf 80}, 1582 (1988)
; M. Doran and J. Jaeckel,  Phys. Rev. D {\bf 66}, 043519 (2002).

\bibitem{Picon:1999} C. A.  Picon, T. Damour and V. Mukhanov, Phys. Lett.
B {\bf 458}, 209  (1999); T. Chiba, T. Okabe and M. Yamaguchi, Phys.
Rev. D {\bf 62}, 023511 (2000).

\bibitem{Caldwell:2002} R. R. Caldwell,  Phys. Lett. B {\bf 545}, 23 (2002) ;
B. McInnes, J. High Energy Phys. {\bf 08}, 029 (2002); S. Nojiri and
S. D. Odintsov, Phys. Lett. B {\bf 562}, 147 (2003);  L. P. Chimento
and R. Lazkoz,  Phys. Rev. Lett. {\bf 91},  211301 (2003); B.
Boisseau, G. Esposito-Farese, D. Polarski, Alexei A. Starobinsky,
Phys. Rev. Lett. {\bf 85}, 2236  (2000); R. Gannouji, D. Polarski,
A. Ranquet, A. A. Starobinsky, JCAP {\bf 0609}, 016 (2006).


\bibitem{Amendola:1993} L. Amendola, Phys. Lett. B {\bf 301}, 175 (1993).

\bibitem{Capozziello:2000} S. Capozziello, G. Lambiase and H. J.Schmidt, Annalen Phys. {\bf9}, 39 (2000).

\bibitem{Capozziello:1999} S. Capozziello, G. Lambiase, Gen. Rel. Grav. {\bf31}, 1005 (1999) .

\bibitem{Sushkov:2009} S. V. Sushkov, Phys. Rev. D {\bf 80}, 103505  (2009).

\bibitem{Gao:2010} C. J. Gao, JCAP {\bf06}, 023 (2010), arXiv: 1002.4035.

\bibitem{Granda:2009} L.N. Granda, arXiv: 0911.3702.

\bibitem{Saridakis:2010} E. N. Saridakis and S. V. Sushkov, Phys. Rev. D {\bf 81}, 083510 (2010), arXiv: 1002.3478.

\bibitem{Chen:2010} S. B. Chen and J. L. Jing, arXiv: 1005.5601.


\bibitem {Gundlach:1994} C. Gundlach, R. H. Price and J. Pullin, Phys. Rev. D {\bf49},  883 (1994).

\bibitem{Schutz:1985}  B. F. Schutz and C. M. Will, Astrophys. J. Lett. {\bf291}, L33 (1985).


\bibitem{Iyer:1987} S. Iyer and C. M. Will, Phys. Rev. D {\bf35}, 3621 (1987); S. Iyer, Phys. Rev. D {\bf35}, 3632
(1987).

\bibitem {Gleiser:2005} R. J. Gleiser and G. Dotti, Phys. Rev. D {\bf72}, 124002
(2005); W. F. Buell and B. A. Shadwick, Am. J. Phys. {\bf63}, 256
(1995).



\end{thebibliography}
\end{document}